\documentclass{article}

\usepackage{newtxtext}
\usepackage{color}
\usepackage{comment}
\usepackage{cite}
\usepackage{hyperref}
\hypersetup{colorlinks=true,linkcolor=blue,citecolor=blue}
\usepackage{amsmath}
\usepackage{amssymb}
\usepackage{amsthm}

\theoremstyle{definition}
\newtheorem{Def}{Definition}[section]

\newtheorem{rem}[Def]{Remark}

\newcommand{\const}{\mathrm{const}.}
\newcommand{\al}{\alpha}
\newcommand{\be}{\beta}
\newcommand{\gam}{\gamma}
\newcommand{\del}{\delta}
\newcommand{\eps}{\epsilon}
\newcommand{\sig}{\sigma}
\newcommand{\ov}[1]{\overline{#1}}
\newcommand{\un}[1]{\underline{#1}}
\newcommand{\bfrac}[2]{\left(\frac{#1}{#2}\right)}
\newcommand{\od}[3][1]{
\ifnum #1=1\frac{\mathrm{d} #2}{\mathrm{d} #3}
\else\frac{\mathrm{d}^{#1} #2}{\mathrm{d} #3^{#1}}
\fi}
\newcommand{\pd}[3][1]{
\ifnum #1=1\frac{\partial #2}{\partial #3}
\else\frac{\partial^{#1} #2}{\partial #3^{#1}}
\fi}
\newcommand{\hd}[4]{D^{#1}_{#2}{#3}\cdot{#4}}

\newcommand{\dlydiff}{delay-differential}
\newcommand{\Dlydiff}{Delay-differential}

\newcommand{\pain}{Painlev\'e}
\newcommand{\lv}{Lotka-Volterra}
\newcommand{\tl}{Toda lattice}
\newcommand{\sg}{sine-Gordon}
\newcommand{\kdv}{KdV equation}
\newcommand{\bsq}{Boussinesq}
\newcommand{\kp}{KP equation}
\usepackage{graphicx}
\usepackage{mathtools}
\newcommand{\fl}{\MoveEqLeft[0]}

\usepackage{authblk}

\begin{document}

\title
{Integrable delay-difference and \dlydiff\ analogues of the KdV, \bsq, and KP equations}
\author[1]{Kenta Nakata
\thanks{kennakaxx@akane.waseda.jp}}
\affil[1]{Department of Pure and Applied Mathematics, School of Fundamental Science and Engineering, Waseda University, 3-4-1 Okubo, Shinjuku-ku, Tokyo 169-8555, Japan}
\date{\today}

\maketitle

\begin{abstract}
    Delay-difference and \dlydiff\ analogues of the KdV and \bsq\ (BSQ) equations are presented.
    Each of them has the $N$-soliton solution and reduces to an already known soliton equation as the delay parameter approaches 0.
    In addition, a \dlydiff\ analogue of the \kp\ is proposed.
    We discuss its $N$-soliton solution and the limit as the delay parameter approaches 0.
    Finally, the relationship between the \dlydiff\ analogues of the KdV, BSQ, and KP equations is clarified.
    Namely, reductions of the delay KP equation yield the delay KdV and delay BSQ equations.
\end{abstract}

\begin{section}{Introduction}
\label{sec_intro}

\Dlydiff\ equations can often model and analyze systems involving feedback effects, such as infectious disease and traffic flow.
In the stream of such studies, exact solutions of \dlydiff\ equations have been obtained and discussed~\cite{nakanishi,hasebe,kanai}.

In recent years, research on integrable \dlydiff\ equations is actively developing.
For examples of ordinary differential equations, integrable \dlydiff\ analogues of \pain\ equations have been introduced by several methods.
Quispel \textit{et al}.~\cite{Quispel} considered a similarity reduction of the \lv\ (LV) equation, then obtained a \dlydiff\ analogue of the first \pain\ equation.
Ramani \textit{et al}.~\cite{Ramani1,Ramani2} considered the integrability test for \dlydiff\ systems, then Grammaticos \textit{et al}.~\cite{Gram} used it and obtained \dlydiff\ analogues of some \pain\ equations.
After these discoveries, various properties of the delay \pain\ equations were revealed from some viewpoints~\cite{Levi,Joshi1,Joshi2,Carstea,Viallet,Halburd,Berntson,Stokes}.

On the other hand, exact solutions of integrable partial \dlydiff\ equations have been studied.
For example, a \dlydiff\ analogue of the two-dimensional \tl\ equation (2DTL) was addressed~\cite{Ablowitz,Tsunematsu}.
To our knowledge, this is the first example of a \textit{delay soliton equation}, which means a \dlydiff\ equation with the $N$-soliton solution.
Subsequently, in our previous study~\cite{Nakata1}, we proposed a systematic method for constructing delay-difference and \dlydiff\ analogues of soliton equations.
As examples of this method, we constructed delay-difference and \dlydiff\ analogues of the LV, \tl\ (TL), and \sg\ (sG) equations with their $N$-soliton solutions.
However, these delay soliton equations are just simple examples, and this method cannot be easily applied to some soliton equations including the \kdv\ (this meaning is explained in Remark \ref{rem_kdv1}).
Thus this method is not yet sufficient to discuss integrable properties of delay soliton equations, such as relationships to delay \pain\ equations.

In section \ref{sec_kdv} of this paper, we first propose integrable delay-difference and \dlydiff\ analogues of the \kdv\ and their $N$-soliton solutions by using the above method~\cite{Nakata1}.
In detail, it is carried out by applying reduction and continuum limits to the discrete \kp~\cite{Hirota-discrete_KP,Miwa-discrete_KP}.
As the delay parameter approaches $0$, this \dlydiff\ analogues of the \kdv\ (the delay \kdv) reduces to the well-known \kdv, and the same is true for the $N$-soliton solution.
We can also obtain a delay \bsq\ (BSQ) equation
by applying a transformation to the delay \kdv.
In section \ref{sec_kp}, we propose a \dlydiff\ analogue of the \kp\ by a transformation from the 2DTL.
The Casorati determinant solution of this delay \kp\ is obtained by using equivalence to the 2DTL.
Similar to the KdV and BSQ cases, the delay \kp\ and its $N$-soliton solution reduce to the \kp\ and its $N$-soliton solution as the delay parameter approaches 0.
Finally, we derive the above delay KdV and delay BSQ equations again by reductions of the delay \kp.

\end{section}

\begin{section}{An integrable delay \kdv}
\label{sec_kdv}

In this section, we construct a delay-difference and \dlydiff\ analogue of the \kdv\ by using the method for constructing delay soliton equations~\cite{Nakata1}.
An overview of this method is given as follows.
\begin{itemize}
    \item [Step-1]
    We consider a reduction of the discrete KP equation.
    The reduction condition needs to include a free parameter $\al$ (such as (\ref{reduction_KdV})).
    Applying this reduction to the discrete \kp, we obtain a discrete equation that depends on $\al$ (such as (\ref{dlydiskdv})).
    This discrete equation turns out to be a delay-difference analogue of a soliton equation.
    \item [Step-2]
    We apply a continuum limit to the above delay-difference soliton equation.
    In this continuum limit, the lattice parameter approaches $0$, and the parameter $\al$ approaches infinity (such as (\ref{dlydiskdv_limit})).
    Then we obtain a \dlydiff\ equation (such as (\ref{dlysdkdv})), which is considered as a \dlydiff\ analogue of a soliton equation.
\end{itemize}
Following the above process, we first construct the \textit{delay discrete} \kdv\ by a reduction of the discrete \kp\ (subsection \ref{subsec_dlydiskdv}).
Then we obtain the \textit{delay semi-discrete} \kdv\ by continuum limit (subsection \ref{subsec_dlysdkdv}).
Finally we obtain the \textit{delay} \kdv\ by continuum limit (subsection \ref{subsec_dlykdv}).

\begin{subsection}{The delay discrete \kdv}
\label{subsec_dlydiskdv}

We start from the discrete \kp~\cite{Hirota-discrete_KP,Miwa-discrete_KP}
\begin{align}
    \label{dkp}
    \fl a(b-c)f_{n+1,m,k}f_{n,m+1,k+1}
    +b(c-a)f_{n,m+1,k}f_{n+1,m,k+1}\nonumber\\
    \fl \hspace{20mm}+c(a-b)f_{n,m,k+1}f_{n+1,m+1,k}
    =0
\end{align}
and its $N$-soliton solution in the Gram determinant form~\cite{Ohta-discrete_KP}
\begin{align}
    \label{dkp_gram}
    \fl f_{n,m,k}
    =\det\left(\delta_{ij}+\frac{\phi_i \psi_j}{p_i-q_j}\right)_{1\leq i,j\leq N}
    =\det\left(\delta_{ij}+\frac{\phi_j \psi_j}{p_i-q_j}\right)_{1\leq i,j\leq N}\,,\\
    \fl \phi_i(n,m,k)
    =\be_i^{-1}(1-ap_i)^{-n}(1-bp_i)^{-m}(1-cp_i)^{-k}\,,\nonumber\\
    \fl \psi_i(n,m,k)
    =\gam_i(1-aq_i)^{n}(1-bq_i)^{m}(1-cq_i)^{k} \,,\nonumber
\end{align}
where $a,b,c,p_i,q_i,\be_i,\gam_i$ are real constants.
Using the function
\begin{align*}
    \Phi_i(n,m,k)
    =\frac{\gam_i}{\be_i}\bfrac{1-aq_i}{1-ap_i}^{n}\bfrac{1-bq_i}{1-bp_i}^{m}\bfrac{1-cq_i}{1-cp_i}^{k}\,,
\end{align*}
we can describe solution (\ref{dkp_gram}) as
\begin{align*}
    f_{n,m,k}
    =\det\left(\delta_{ij}+\frac{\Phi_j}{p_i-q_j}\right)_{1\leq i,j\leq N}\,.
\end{align*}
If we ignore the constant doubling of $\be_i$ and $\gam_i$, this solution can be rewritten as follows:
\begin{align}
    \label{dkp_nsol}
    \fl f_{n,m,k}
    =\det\left(\delta_{ij}+\frac{\Phi_j}{p_i-q_j}\right)_{1\leq i,j\leq N}
    =\sum_{I\subset\{1,\ldots,N\}}
    \prod_{i\in I}\Phi_i
    \prod_{i<j,\ i,j\in I}\frac{(p_i-p_j)(q_i-q_j)}{(p_i-q_j)(q_i-p_j)}\,.
\end{align}
For convenience, we apply the transformation $n'=2(n+m),\ \ m'=2(n-m)$ to the discrete \kp\ (\ref{dkp}) and its solution (\ref{dkp_nsol}), then rewrite $n',m'$ by $n,m$ again.
Through the transformation, equation (\ref{dkp}) is transformed into
\begin{align}
    \label{dkptrans}
    \fl a(b-c)f_{n+2,m+2,k}f_{n+2,m-2,k+1}
    +b(c-a)f_{n+2,m-2,k}f_{n+2,m+2,k+1}\nonumber\\
    \fl \hspace{20mm}+c(a-b)f_{n,m,k+1}f_{n+4,m,k}
    =0\,.
\end{align}
Replacing $p_i$ and $q_i$ by $p_i/a$ and $q_i/a$ respectively, and setting $\mu_i=\gam_i/\be_i$, solution (\ref{dkp_nsol}) is transformed into
\begin{align}
    \label{dkptrans_sol}
    \fl f_{n,m,k}
    =\det\left(\delta_{ij}+\frac{\Phi_j}{p_i-q_j}\right)_{1\leq i,j\leq N}
    =\sum_{I\subset\{1,\ldots,N\}}
    \prod_{i\in I}\Phi_i
    \prod_{i<j,\ i,j\in I}\frac{(p_i-p_j)(q_i-q_j)}{(p_i-q_j)(q_i-p_j)}\,,\\
    \fl\Phi_i
    =\mu_i
    \left(\frac{1-q_i}{1-p_i}\frac{a-bq_i}{a-bp_i}\right)^{n/4}
    \left(\frac{1-q_i}{1-p_i}\frac{a-bp_i}{a-bq_i}\right)^{m/4}
    \bfrac{a-cq_i}{a-cp_i}^{k}\,.\nonumber
\end{align}

Now, we apply the reduction condition
\begin{align}
    \label{reduction_KdV}
    f_{n,m,k+1}=f_{n-2-2\nu,m-2-2\al,k}
\end{align}
to the transformed discrete \kp\ (\ref{dkptrans}) and its solution (\ref{dkptrans_sol}), where the parameters $\al$ and $\nu$ are fixed values considered as the delays.
Setting $c=-a$ and $\eps\del=(b-a)/(b+a)$, the transformed discrete \kp\ (\ref{dkptrans}) reduces to the following bilinear equation
\begin{align}
    \label{dlydiskdv}
    \eps\del f_{n+3+\nu}^{m+1+\al}f_{n-3-\nu}^{m-1-\al}
    +f_{n+1+\nu}^{m+3+\al}f_{n-1-\nu}^{m-3-\al}
    -(1+\eps\del)f_{n+1+\nu}^{m-1+\al}f_{n-1-\nu}^{m+1-\al}
    =0\,,
\end{align}
where $f_{n}^{m}\equiv f_{n,m,k}$ (the dependence of $k$ is omitted).
The variables $n$ and $m$ are considered to be the discrete space and time variables respectively.
We can rewrite the bilinear equation (\ref{dlydiskdv}) as follows by using Hirota's D-operators:
\begin{align}
    \label{dlydiskdv2}
    \fl(2\sinh\left(D_n+D_m+\nu D_n+\al D_m\right)\sinh(2D_m)\nonumber\\
    \fl \hspace{20mm}+2\eps\del\sinh(D_n+D_m)\sinh\left(2D_n+\nu D_n+\al D_m\right))
    f_{n}^{m}\cdot f_{n}^{m}=0\,.
\end{align}
Here Hirota's D-operators are defined by
\begin{align*}
    \fl\hd{l}{t}{g(t)}{h(t)}
    =\left(\pd{}{t}-\pd{}{s}\right)^{l}g(t)h(s)\Bigr \vert_{s=t}\,,\qquad
    e^{rD_m}g_{m}\cdot h_{m}
    =g_{m+r}h_{m-r}\,.
\end{align*}
We can consider equation (\ref{dlydiskdv}) (or (\ref{dlydiskdv2})) is the bilinear form of the delay-difference analogue of the \kdv.
By putting $\nu=\al=0$ and $\eps=1$, we obtain the bilinear form of the discrete \kdv, which is called the difference-difference \kdv~\cite{Hirota-discrete_KdV,Maruno-discrete_soliton_equations}.
We call equation (\ref{dlydiskdv}) the bilinear form of the delay discrete \kdv.

Next we consider the reduction of the $N$-soliton solution.
To realize the reduction condition (\ref{reduction_KdV}), it is sufficient to impose the constraint
\begin{align*}
    1
    =\left(\frac{1-q_i}{1-p_i}\frac{a-bq_i}{a-bp_i}\right)^{(1+\nu)/2}
    \left(\frac{1-q_i}{1-p_i}\frac{a-bp_i}{a-bq_i}\right)^{(1+\al)/2}
    \bfrac{a-cq_i}{a-cp_i}
\end{align*}
on the $N$-soliton solution (\ref{dkptrans_sol}).
The reason is as follows.
By this constraint, we can obtain the relation
\begin{align*}
    &\Phi_i(n,m,k+1)\\
    =&\mu_i
    \left(\frac{1-q_i}{1-p_i}\frac{a-bq_i}{a-bp_i}\right)^{n/4}
    \left(\frac{1-q_i}{1-p_i}\frac{a-bp_i}{a-bq_i}\right)^{m/4}
    \bfrac{a-cq_i}{a-cp_i}^{k+1}\\
    =&\mu_i
    \left(\frac{1-q_i}{1-p_i}\frac{a-bq_i}{a-bp_i}\right)^{(n-2-2\nu)/4}
    \left(\frac{1-q_i}{1-p_i}\frac{a-bp_i}{a-bq_i}\right)^{(m-2-2\al)/4}
    \bfrac{a-cq_i}{a-cp_i}^{k}\\
    =&\Phi_i(n-2-2\nu,m-2-2\al,k)\,.
\end{align*}
Solution (\ref{dkptrans_sol}) with this relation $\Phi_i(n,m,k+1)=\Phi_i(n-2-2\nu,m-2-2\al,k)$ satisfies the condition (\ref{reduction_KdV}): $f_{n,m,k+1}=f_{n-2-2\nu,m-2-2\al,k}$, because $f$ is a function of $\Phi_i(n,m,k)\ (1\leq i\leq N)$.
Now, using $c=-a$, $\eps\del=(b-a)/(b+a)$ and setting $k=0$, we obtain the $N$-soliton solution of the delay discrete \kdv\ (\ref{dlydiskdv}) as follows:
\begin{align}
    \label{dlydiskdv_sol}
    \fl f_{n}^{m}
    =\det\left(\delta_{ij}+\frac{\Phi_j}{p_i-q_j}\right)_{1\leq i,j\leq N}
    =\sum_{I\subset\{1,\ldots,N\}}
    \prod_{i\in I}\Phi_i
    \prod_{i<j,\ i,j\in I}\frac{(p_i-p_j)(q_i-q_j)}{(p_i-q_j)(q_i-p_j)}\,,\\
    \fl\Phi_i
    =\mu_i\left(\frac{1-q_i}{1-p_i}\frac{1-q_i-\eps\del(1+q_i)}{1-p_i-\eps\del(1+p_i)}\right)^{n/4}
    \left(\frac{1-q_i}{1-p_i}\frac{1-p_i-\eps\del(1+p_i)}{1-q_i-\eps\del(1+q_i)}\right)^{m/4}\,,\nonumber\\
    \fl1
    =\left(\frac{1-q_i}{1-p_i}\frac{1-q_i-\eps\del(1+q_i)}{1-p_i-\eps\del(1+p_i)}\right)^{(1+\nu)/2}
    \left(\frac{1-q_i}{1-p_i}\frac{1-p_i-\eps\del(1+p_i)}{1-q_i-\eps\del(1+q_i)}\right)^{(1+\al)/2}
    \bfrac{1+q_i}{1+p_i}\,.\nonumber
\end{align}
The solution (\ref{dlydiskdv_sol}) in the case of $\nu=\al=0$ and $\eps=1$ is equivalent to the $N$-soliton solution of the discrete \kdv~\cite{Hirota-discrete_KdV}.

Now, let us derive the nonlinear form of the delay discrete \kdv.
We consider the dependent variable transformation
\begin{align}
    \label{dlydiskdv_trans}
    w_{n}^{m}
    =\frac{f_{n+2+(\nu/2)}^{m+(\al/2)}f_{n-2-(\nu/2)}^{m-(\al/2)}}
    {f_{n+(\nu/2)}^{m+2+(\al/2)}f_{n-(\nu/2)}^{m-2-(\al/2)}}\,.
\end{align}
Through this transformation, the bilinear equation (\ref{dlydiskdv}) is transformed into
\begin{align}
    \label{dlydiskdv_nl}
    \fl\frac{
     w_{n+1+\nu}^{m-1+\al}
     w_{n-1-\nu}^{m+1-\al}}
    {w_{n+1+\nu}^{m+3+\al}
     w_{n-1-\nu}^{m-3-\al}}
    =\frac{
     \left(1+\eps\del w_{n+3+\nu}^{m+1+\al}w_{n+1}^{m-1}\right)
     \left(1+\eps\del w_{n-1}^{m+1}w_{n-3-\nu}^{m-1-\al}\right)}
    {\left(1+\eps\del w_{n+1+\nu}^{m+3+\al}w_{n-1}^{m+1}\right)
     \left(1+\eps\del w_{n+1}^{m-1}w_{n-1-\nu}^{m-3-\al}\right)}\,.
\end{align}
This is the nonlinear form of the delay discrete \kdv.
When $\al=\nu=0$ and $\eps=1$, (\ref{dlydiskdv_nl}) is the division of the following two equations:
\begin{align}
    \label{diskdv_nl1}
    \fl \frac{w_{n+1}^{m-1}}{w_{n+1}^{m+3}}
    =\frac{1+\del w_{n+3}^{m+1}w_{n+1}^{m-1}}{1+\del w_{n+1}^{m+3}w_{n-1}^{m+1}}\,,\\
    \label{diskdv_nl2}
    \fl \frac{w_{n-1}^{m-3}}{w_{n-1}^{m+1}}
    =\frac{1+\del w_{n+1}^{m-1}w_{n-1}^{m-3}}{1+\del w_{n-1}^{m+1}w_{n-3}^{m-1}}\,.
\end{align}
Both (\ref{diskdv_nl1}) and (\ref{diskdv_nl2}) are equivalent to the discrete \kdv~\cite{Hirota-discrete_KdV}
\begin{align*}
    \frac{1}{w_{n}^{m+2}}-\frac{1}{w_{n}^{m-2}}=\del(w_{n+2}^{m}-w_{n-2}^{m})\,.
\end{align*}

\begin{rem}
\label{rem_diskdv}
Transformation (\ref{dlydiskdv_trans}) is obtained by an analogy to the case of the discrete \kdv.

First, let us remind of the case of the discrete \kdv.
The bilinear form is
\begin{align*}
    \del f_{n+3}^{m+1}f_{n-3}^{m-1}
    +f_{n+1}^{m+3}f_{n-1}^{m-3}
    -(1+\del)f_{n+1}^{m-1}f_{n-1}^{m+1}
    =0\,,
\end{align*}
which is rewritten as
\begin{align*}
    (1+\del)\frac{f_{n+1}^{m-1}f_{n-1}^{m+1}}
    {f_{n+1+\nu}^{m+3}f_{n-1}^{m-3}}
    =1
    +\del\frac{f_{n+3}^{m+1}f_{n-3}^{m-1}}
    {f_{n+1+\nu}^{m+3}f_{n-1}^{m-3}}\,.
\end{align*}
Putting $w_{n}^{m}=(f_{n+2}^{m}f_{n-2}^{m})/(f_{n}^{m+2}f_{n}^{m-2})$, we obtain
\begin{align*}
    (1+\del)\frac{f_{n+1}^{m-1}f_{n-1}^{m+1}}
    {f_{n+1+\nu}^{m+3}f_{n-1}^{m-3}}
    =1
    +\del w_{n+1}^{m+1}w_{n-1}^{m-1}\,.
\end{align*}
By shifting and combining this equation, we obtain the nonlinear form (\ref{diskdv_nl1}) (or (\ref{diskdv_nl2})).

Now, inspired by this derivation, we rewrite delayed bilinear equation (\ref{dlydiskdv}) as
\begin{align*}
    (1+\eps\del)\frac{f_{n+1+\nu}^{m-1+\al}f_{n-1-\nu}^{m+1-\al}}
    {f_{n+1+\nu}^{m+3+\al}f_{n-1-\nu}^{m-3-\al}}
    =1
    +\eps\del\frac{f_{n+3+\nu}^{m+1+\al}f_{n-3-\nu}^{m-1-\al}}
    {f_{n+1+\nu}^{m+3+\al}f_{n-1-\nu}^{m-3-\al}}\,.
\end{align*}
Putting $w_{n}^{m}$ as (\ref{dlydiskdv_trans}), we obtain
\begin{align*}
    (1+\eps\del)\frac{f_{n+1+\nu}^{m-1+\al}f_{n-1-\nu}^{m+1-\al}}
    {f_{n+1+\nu}^{m+3+\al}f_{n-1-\nu}^{m-3-\al}}
    =1
    +\eps\del w_{n+1+(\nu/2)}^{m+1+(\al/2)}w_{n-1-(\nu/2)}^{m-1-(\al/2)}\,.
\end{align*}
By shifting and combining this equation, we obtain the nonlinear form (\ref{dlydiskdv_nl}).
\end{rem}

\begin{rem}
Let us describe the delay discrete KdV equation (\ref{dlydiskdv_nl}) in other independent variables.
We apply the transformation
\begin{align*}
    n'=\frac{1}{4}(n+m),\qquad
    m'=\frac{1}{4}(-n+m),\qquad
    \nu'=\frac{1}{4}(\nu+\al),\qquad
    \al'=\frac{1}{4}(-\nu+\al)
\end{align*}
to equation (\ref{dlydiskdv_nl}), then rewrite $n',m',\al',\nu'$ by $n,m,\al,\nu$ again.
Through this transformation, equation (\ref{dlydiskdv_nl}) is transformed into
\begin{align*}
    \fl\frac{
     w_{n+\nu}^{m+\al}
     w_{n-\nu}^{m+1-\al}}
    {w_{n+1+\nu}^{m+1+\al}
     w_{n-1-\nu}^{m-\al}}
    =\frac{
     \left(1+\del w_{n+1+\nu}^{m+\al}w_{n}^{m}\right)
     \left(1+\del w_{n}^{m+1}w_{n-1-\nu}^{m+1-\al}\right)}
    {\left(1+\del w_{n+1+\nu}^{m+1+\al}w_{n}^{m+1}\right)
     \left(1+\del w_{n}^{m}w_{n-1-\nu}^{m-\al}\right)}\,.
\end{align*}
When $\al=\nu=0$, this equation is the division of the following two equations:
\begin{align*}
    \fl \frac{w_{n+1}^{m+1}}{w_{n}^{m}}
    =\frac{1+\del w_{n+1}^{m+1}w_{n}^{m+1}}{1+\del w_{n+1}^{m}w_{n}^{m}}\,,\qquad
    \frac{w_{n}^{m+1}}{w_{n-1}^{m}}
    =\frac{1+\del w_{n}^{m+1}w_{n-1}^{m+1}}{1+\del w_{n}^{m}w_{n-1}^{m}}\,,
\end{align*}
which are equivalent to the discrete \kdv
\begin{align*}
    \frac{1}{w_{n+1}^{m+1}}-\frac{1}{w_{n}^{m}}=\del(w_{n+1}^{m}-w_{n}^{m+1})\,.
\end{align*}
\end{rem}

\end{subsection}

\begin{subsection}{The delay semi-discrete \kdv}
\label{subsec_dlysdkdv}

We propose the following continuum limit to the delay discrete \kdv\ (\ref{dlydiskdv}) and its $N$-soliton solution (\ref{dlydiskdv_sol}):
\begin{align}
    \label{dlydiskdv_limit}
    \del\to0\,,\qquad
    \frac{m\del}{4}=t\,,\qquad 
    \frac{\al\del}{4}=\tau=\const\,,
\end{align}
where $t$ is the continuous time variable, and $\tau$ is a fixed value considered as the time-delay.
Applying the continuum limit (\ref{dlydiskdv_limit}) to (\ref{dlydiskdv}) and (\ref{dlydiskdv_sol}), we obtain the bilinear equation
\begin{align}
    \label{dlysdkdv}
    \fl \hd{}{t}{f_{n+1+\nu}(t+\tau)}{f_{n-1-\nu}(t-\tau)}
    +\eps f_{n+3+\nu}(t+\tau)f_{n-3-\nu}(t-\tau)\nonumber\\
    \fl \hspace{20mm}-\eps f_{n+1+\nu}(t+\tau)f_{n-1-\nu}(t-\tau)
    =0
\end{align}
and its $N$-soliton solution
\begin{align}
    \label{dlysdkdv_sol}
    \fl f_{n}(t)
    =\det\left(\delta_{ij}+\frac{\Phi_j}{p_i-q_j}\right)_{1\leq i,j\leq N}
    =\sum_{I\subset\{1,\ldots,N\}}
    \prod_{i\in I}\Phi_i
    \prod_{i<j,\ i,j\in I}\frac{(p_i-p_j)(q_i-q_j)}{(p_i-q_j)(q_i-p_j)}\,,\\
    \fl\Phi_i
    =\mu_i\bfrac{1-q_i}{1-p_i}^{n/2}\exp\left\{\eps\left(-\frac{1+p_i}{1-p_i}+\frac{1+q_i}{1-q_i}\right)t\right\}\,,\nonumber\\
    \fl1
    =\bfrac{1-q_i}{1-p_i}^{1+\nu}\exp\left\{2\tau\eps\left(-\frac{1+p_i}{1-p_i}+\frac{1+q_i}{1-q_i}\right)\right\}\bfrac{1+q_i}{1+p_i}\,.\nonumber
\end{align}
Bilinear equation (\ref{dlysdkdv}) is rewritten as
\begin{align}
    \label{dlysdkdv2}
    \fl\left(\sinh\left(D_n+\nu D_n+\tau D_t\right)D_t
    +2\eps\sinh(D_n)\sinh\left(2D_n+\nu D_n+\tau D_t\right)\right)
    f_{n}(t)\cdot f_{n}(t)\nonumber\\
    \fl \hspace{20mm}=0\,.
\end{align}
We can claim equation (\ref{dlysdkdv}) (or (\ref{dlysdkdv2})) is the bilinear form of the delay semi-discrete \kdv.
By putting $\nu=\tau=0$ and $\eps=1$, we can obtain the bilinear form of the semi-discrete \kdv, which is called the differential-difference \kdv~\cite{Hirota-discrete_KdV}.
In addition, the solution (\ref{dlysdkdv_sol}) in the case of $\nu=\tau=0$ and $\eps=1$ is equivalent to the $N$-soliton solution of the semi-discrete \kdv~\cite{Hirota-discrete_KdV}.

\begin{rem}
The delay LV equation~\cite{Nakata1} is given by
\begin{align*}
    \hd{}{t}{f_{n}(t+\tau)}{f_{n-1}(t-\tau)}
    -f_{n+1}(t+\tau)f_{n-2}(t-\tau)
    +f_{n}(t+\tau)f_{n-1}(t-\tau)
    =0\,.
\end{align*}
This equation can be derived by applying the transformation
\begin{align*}
    n\rightarrow 2n\,,\quad
    t\rightarrow -t\,,\quad
    \tau\rightarrow -\tau\,,\quad
    \nu=0\,,\quad
    \eps=1
\end{align*}
to the delay semi-discrete \kdv\ (\ref{dlysdkdv}).
\end{rem}

\begin{rem}
The semi-discrete KP equation~\cite{Li-semi_discrete_KP} is given by
\begin{align*}
    D_{t}f_{n+2}^{k}(t)\cdot f_{n}^{k+1}(t)
    +\eps(f_{n+2}^{k}(t)f_{n}^{k+1}(t)-f_{n+2}^{k+1}(t)f_{n}^{k}(t))
    =0\,.
\end{align*}
Applying the reduction condition $f_{n}^{k+1}(t)=f_{n+4+2\nu}^{k}(t+2\tau)$ to this equation, we can derive the delay semi-discrete \kdv\ (\ref{dlysdkdv}).
\end{rem}

Now, let us derive the nonlinear form of the delay semi-discrete \kdv.
We consider the dependent variable transformation
\begin{align}
    \label{dlysdkdv_trans}
    w_{n}(t)
    =\frac{f_{n+2+(\nu/2)}(t+\tau/2)f_{n-2-(\nu/2)}(t-\tau/2)}
    {f_{n+(\nu/2)}(t+\tau/2)f_{n-(\nu/2)}(t-\tau/2)}\,.
\end{align}
Through this transformation, the bilinear equation (\ref{dlysdkdv}) is transformed into
\begin{align}
    \label{dlysdkdv_nl}
    \od{}{t}\log\frac{w_{n+1+\nu}(t+\tau)}{w_{n-1-\nu}(t-\tau)}
    &=     -\eps w_{n+3+\nu}(t+\tau)\ w_{n+1}(t)
           -\eps w_{n-1}(t)\ w_{n-3-\nu}(t-\tau)\nonumber\\
    &\quad +\eps w_{n+1+\nu}(t+\tau)\ w_{n-1}(t)
           +\eps w_{n+1}(t)\ w_{n-1-\nu}(t-\tau)\,.
\end{align}
This is the nonlinear form of the delay semi-discrete \kdv.
When $\tau=\nu=0$ and $\eps=1$, (\ref{dlysdkdv_nl}) is the subtraction of the following two equations:
\begin{align}
    \label{sdkdv_nl1}
    \fl \od{}{t}\log(w_{n+1}(t))
    =w_{n+1}(t)(w_{n-1}(t)-w_{n+3}(t))\,,\\
    \label{sdkdv_nl2}
    \fl \od{}{t}\log(w_{n-1}(t))
    =w_{n-1}(t)(w_{n-3}(t)-w_{n+1}(t))\,.
\end{align}
Both (\ref{sdkdv_nl1}) and (\ref{sdkdv_nl2}) are equivalent to the semi-discrete \kdv~\cite{Hirota-discrete_KdV}
\begin{align*}
    \od{}{t}\frac{1}{w_{n}(t)}
    =w_{n+2}(t)-w_{n-2}(t)\,.
\end{align*}

\begin{rem}
Similar to Remark \ref{rem_diskdv}, transformation (\ref{dlysdkdv_trans}) is obtained by an analogy to the case of the semi-discrete \kdv.

First, let us remind of the case of the semi-discrete \kdv.
The bilinear form is
\begin{align*}
    \hd{}{t}{f_{n+1}(t)}{f_{n-1}(t)}
    +f_{n+3}(t)f_{n-3}(t)
    -f_{n+1}(t)f_{n-1}(t)
    =0\,,
\end{align*}
which is rewritten as
\begin{align*}
    \od{}{t}\log\frac{f_{n+1}(t)}{f_{n-1}(t)}
    =1
    -\frac{f_{n+3}(t)f_{n-3}(t)}{f_{n+1}(t)f_{n-1}(t)}\,.
\end{align*}
Putting $w_{n}(t)=(f_{n+2}(t)f_{n-2}(t))/(f_{n}(t)f_{n}(t))$, we obtain
\begin{align*}
    \od{}{t}\log\frac{f_{n+1}(t)}{f_{n-1}(t)}
    =1
    -w_{n+1}(t)w_{n-1}(t)\,.
\end{align*}
By shifting and combining this equation, we obtain the nonlinear form (\ref{sdkdv_nl1}) (or (\ref{sdkdv_nl2})).

Now, inspired by this derivation, we rewrite delayed bilinear equation (\ref{dlysdkdv}) as
\begin{align*}
    \od{}{t}\log\frac{f_{n+1+\nu}(t+\tau)}{f_{n-1-\nu}(t-\tau)}
    =\eps
    -\eps\frac{f_{n+3+\nu}(t+\tau)f_{n-3-\nu}(t-\tau)}
    {f_{n+1+\nu}(t+\tau)f_{n-1-\nu}(t-\tau)}\,.
\end{align*}
Putting $w_{n}(t)$ as (\ref{dlysdkdv_trans}), we obtain
\begin{align*}
    \od{}{t}\log\frac{f_{n+1+\nu}(t+\tau)}{f_{n-1-\nu}(t-\tau)}
    =\eps
    -\eps w_{n+1+(\nu/2)}(t+\tau/2)w_{n-1-(\nu/2)}(t-\tau/2)\,.
\end{align*}
By shifting and combining this equation, we obtain the nonlinear form (\ref{dlysdkdv_nl}).
\end{rem}

\end{subsection}

\begin{subsection}{The delay \kdv}
\label{subsec_dlykdv}

We propose the following continuum limit and transformation to the delay semi-discrete \kdv\ (\ref{dlysdkdv2}) and its $N$-soliton solution (\ref{dlysdkdv_sol}):
\begin{align}
    \label{dlysdkdv_limit}
    \begin{aligned}
    \fl \eps\to0\,,\qquad
    n=\frac{a_2x-a_1s}{a_2\xi-a_1\sig}\,,\qquad
    t=\frac{3}{2}\ \frac{-\sig x+\xi s}{a_2\xi-a_1\sig}\,,\\
    \fl \nu=\frac{a_2(-\xi+a_3\eps)+a_1\sig}{a_2\xi-a_1\sig}\,,\qquad
    \tau=\frac{3}{2}\ \frac{-a_3\sig\eps}{a_2\xi-a_1\sig}\,,
    \end{aligned}
\end{align}
where $x,s,\xi$, and $\sig$ are considered to be the continuous space variable, continuous time variable, space-delay, and time-delay respectively.
We assume the parameters $a_1,a_2,a_3,\xi,\sig$ are real constants.
The relations (\ref{dlysdkdv_limit}) lead to the following ones:
\begin{align}
    \label{dlykdv_dop}
    D_n=\xi D_x+\sig D_s\,,\quad
    D_t=\frac{2}{3}(a_1 D_x+a_2 D_s)\,,\quad
    \nu D_n+\tau D_t=-\xi D_x-\sig D_s+a_3\eps D_x\,.
\end{align}
Let us apply these relations (\ref{dlykdv_dop}) and $\eps\to0$ to the delay semi-discrete \kdv\ (\ref{dlysdkdv2}). Then we find the order of $\eps^0$ vanishes, and obtain the following equation as the order of $\eps^1$:
\begin{align}
    \label{dlykdv2}
    \left((a_1D_x+a_2D_s)(a_3D_x)
    +3\sinh^2(\xi D_x+\sig D_s)\right)
    f(x,s)\cdot f(x,s)=0\,.
\end{align}
This bilinear equation is equivalent to
\begin{align}
    \label{dlykdv1}
    \fl a_1a_3\hd{2}{x}{f(x,s)}{f(x,s)}
    +a_2a_3D_s\hd{}{x}{f(x,s)}{f(x,s)}\nonumber\\
    \fl \hspace{20mm}+\frac{3}{2}\left\{f(x+2\xi,s+2\sig)f(x-2\xi,s-2\sig)
    -f(x,s)f(x,s)\right\}
    =0\,.
\end{align}

Before we calculate the limit of the $N$-soliton solution (\ref{dlysdkdv_sol}), we replace $p_i$ and $q_i$ by $1-(2\eps)/q_i$ and $1-(2\eps)/p_i$ respectively.
Then applying the continuum limit (\ref{dlysdkdv_limit}) to (\ref{dlysdkdv_sol}), we obtain the $N$-soliton solution of the bilinear equation (\ref{dlykdv1}) as follows:
\begin{align}
    \label{dlykdv_sol}
    \fl f(x,s)
    =\det\left(\delta_{ij}+\frac{\Phi_j}{p_i-q_j}\right)_{1\leq i,j\leq N}
    =\sum_{I\subset\{1,\ldots,N\}}
    \prod_{i\in I}\Phi_i
    \prod_{i<j,\ i,j\in I}\frac{(p_i-p_j)(q_i-q_j)}{(p_i-q_j)(q_i-p_j)}\,,\\
    \fl\Phi_i
    =\mu_i\exp\left\{
    \frac{1}{2}\ \frac{a_2x-a_1s}{a_2\xi-a_1\sig}\log\bfrac{q_i}{p_i}
    +\frac{3}{2}\ \frac{-\sig x+\xi s}{a_2\xi-a_1\sig}(q_i-p_i)
    \right\}\,,\nonumber\\
    \fl0
    =a_2a_3\log\bfrac{q_i}{p_i}
    -3a_3\sig(q_i-p_i)
    -(a_2\xi-a_1\sig)\left(\frac{1}{q_i}-\frac{1}{p_i}\right)\,.\nonumber
\end{align}
We claim that equation (\ref{dlykdv1}) (or (\ref{dlykdv2})) is the bilinear form of the delay \kdv.
It is because equation (\ref{dlykdv2}) has a continuum limit to the \kdv.
To check this, we set the following conditions on the parameters:
\begin{align}
    \label{dlykdv_para}
    a_1=3\xi\,,\qquad
    a_2=4\xi^3\,,\qquad
    a_3=-\xi\,,\qquad
    \sig=0\,.
\end{align}
Calculating the limit of (\ref{dlykdv2}) as $\xi\to0$, we obtain the bilinear form of the \kdv~\cite{Hirota-KdV,Hirota-direct}
\begin{align}
    \label{kdv}
    \left(-4D_sD_x+D^4_x\right)
    f(x,s)\cdot f(x,s)=0\,.
\end{align}
As for the $N$-soliton solution (\ref{dlykdv_sol}), we first replace $p_i$ and $q_i$ as follows:
\begin{align}
    \label{dlykdv_rep}
    p_i\rightarrow\frac{1}{1+2\xi p_i}\,,\qquad
    q_i\rightarrow\frac{1}{1+2\xi q_i}\,.
\end{align}
Then applying the limit $\xi\to0$ to (\ref{dlykdv_sol}) under the conditions (\ref{dlykdv_para}), we can obtain the $N$-soliton solution of the \kdv~\cite{Hirota-KdV,Hirota-direct}
\begin{align}
    \label{kdv_sol}
    \fl f(x,s)
    =\det\left(\delta_{ij}+\frac{\Phi_j}{p_i+p_j}\right)_{1\leq i,j\leq N}
    =\sum_{I\subset\{1,\ldots,N\}}
    \prod_{i\in I}\Phi_i
    \prod_{i<j,\ i,j\in I}\frac{(p_i-p_j)^2}{(p_i+p_j)^2}\,,\\
    \fl\Phi_i
    =\mu_ie^{2p_ix+2p_i^3s}\,.\nonumber
\end{align}

\begin{rem}
Under (\ref{dlykdv_para}) and (\ref{dlykdv_rep}), the last equation of (\ref{dlykdv_sol}), which is the dispersion relation, leads to the following relations:
\begin{align*}
    p_i+q_i=O(\xi)\,,\qquad
    p_i+q_i=\frac{4\xi p_i^2}{3}+O(\xi^2)\,.
\end{align*}
Using these relations, we can omit $q_i$ from (\ref{dlykdv_sol}) in the small limit of $\xi$.
Then we can obtain the $N$-soliton solution of the \kdv\ (\ref{kdv_sol}).
\end{rem}

\begin{rem}
\label{rem_kdv1}
According to \cite{Nakata1} and subsection \ref{subsec_dlysdkdv}, the continuum limits to derive the delay LV, delay TL, delay sG, and delay semi-discrete KdV equations were proposed by the simple idea explained below.
While, the continuum limit to derive the delay \kdv\ (\ref{dlysdkdv_limit}) cannot be proposed by this idea, but is proposed heuristically.
We explain about this.

First, for example we remind of the bilinear form of the discrete \kdv:
\begin{align*}
    \del f_{n+3}^{m+1}f_{n-3}^{m-1}
    +f_{n+1}^{m+3}f_{n-1}^{m-3}
    -(1+\del)f_{n+1}^{m-1}f_{n-1}^{m+1}
    =0\,.
\end{align*}
Applying the continuum limit $\del\to0,\ m\del/4=t$, we obtain the bilinear form of the semi-discrete \kdv:
\begin{align*}
    \hd{}{t}{f_{n+1}(t)}{f_{n-1}(t)}
    +f_{n+3}(t)f_{n-3}(t)
    -f_{n+1}(t)f_{n-1}(t)
    =0\,.
\end{align*}
In subsection \ref{subsec_dlysdkdv}, we showed the delay version of this derivation.
The delay discrete \kdv\ (\ref{dlydiskdv}) reduces to the delay semi-discrete \kdv\ (\ref{dlysdkdv}) in the continuum limit (\ref{dlydiskdv_limit}): $\del\to0,\ m\del/4=t,\ \al\del/4=\tau$.
As we can see from this example, we just need to set the relation in which variables $m,t$ are replaced by delay parameters $\al,\tau$ when we consider the delay version.
This idea can also be applied to the delay LV, delay TL, and delay sG equations, and the continuum limits to derive them were easily proposed.

On the other hand, we cannot apply this idea to the delay \kdv.
Actually the above bilinear form of the semi-discrete \kdv\ reduces to that of the \kdv\ (\ref{kdv}) in the continuum limit $\eps\to0,\ x=-(n\eps)/2+2t\eps,\ s=4t\eps^3/3$.
However, as for the delay version, the delay semi-discrete \kdv\ (\ref{dlysdkdv}) does not reduce to a good equation even if we set the relations $\xi=-(\nu\eps)/2+2\tau\eps,\ \sig=4\tau\eps^3/3$.

Now, we need an alternative idea.
Let us focus on the bilinear form (\ref{dlysdkdv2}) described by D-operators.
This time, the second term on the left-hand side is $\mathrm{O}(\eps^1)$.
Hence, if we assume that $D_n+\nu D_n+\tau D_t$ is $\mathrm{O}(\eps^1)$ and $D_t$ is $\mathrm{O}(\eps^0)$, we expect that the first and second terms balance when $\eps\to0$ and a good equation is obtained.
Therefore, we reached to set the relations (\ref{dlykdv_dop}).
The delay \kdv\ was obtained by the heuristic idea which is not parallel to the non-delay version.
\end{rem}

We present the nonlinear form of the delay \kdv\ via the dependent variable transformation
\begin{align}
    \label{dlykdv_trans}
    1+b_1U(x,s)=\frac{f(x+2\xi,s+2\sig)f(x-2\xi,s-2\sig)}{f(x,s)f(x,s)}\,,
\end{align}
where $b_1$ is a real constant.
By using this, bilinear equation (\ref{dlykdv1}) is transformed into the following \dlydiff\ equation:
\begin{align}
    \label{dlykdv_nl}
    \fl a_1a_3\pd[2]{}{x}\log(1+b_1U(x,s))
    +a_2a_3\frac{\partial^2}{\partial x\partial s}\log(1+b_1U(x,s))\nonumber\\
    \fl \hspace{20mm}+\frac{3b_1}{4}\left\{U(x+2\xi,s+2\sig)+U(x-2\xi,s-2\sig)
    -2U(x,s)\right\}
    =0\,.
\end{align}
(\ref{dlykdv_nl}) is the nonlinear form of the delay \kdv.
Under the conditions (\ref{dlykdv_para}) and $b_1=2\xi^2$, the limit of (\ref{dlykdv_nl}) as $\xi\to0$ is the nonlinear form of the \kdv~\cite{Hirota-KdV,Hirota-direct}
\begin{align*}
    -4\pd{}{s}U(x,s)+6U(x,s)\pd{}{x}U(x,s)+\pd[3]{}{x}U(x,s)=0\,.
\end{align*}

\begin{rem}
\label{rem_kdv2}
Transformation (\ref{dlykdv_trans}) is understood as follows.
First, we can rewrite the bilinear equation (\ref{dlykdv1}) as
\begin{align*}
    2a_1a_3(\log f(x,s))_{xx}
    +2a_2a_3(\log f(x,s))_{xs}
    +\frac{3}{2}\left(\frac{f(x+2\xi,s+2\sig)f(x-2\xi,s-2\sig)}{f(x,s)f(x,s)}
    -1\right)
    =0\,.
\end{align*}
To write the third term in the dependent variable $U(x,s)$, we can put (\ref{dlykdv_trans}).
Then we obtain
\begin{align*}
    2a_1a_3(\log f(x,s))_{xx}
    +2a_2a_3(\log f(x,s))_{xs}
    +\frac{3b_1}{2}U(x,s)
    =0\,.
\end{align*}
By shifting and combining this equation, we obtain the nonlinear form (\ref{dlykdv_nl}).

Note that when $b_1=2\xi^2$ and $\xi\to0$, (\ref{dlykdv_trans}) becomes
\begin{align*}
    U(x,s)=2(\log f(x,s))_{xx}\,,
\end{align*}
which is the transformation between the bilinear and nonlinear KdV equations.
\end{rem}

\end{subsection}

\begin{subsection}{The delay BSQ equation}
\label{subsec_dlybsq}

In fact, the delay \kdv\ can be considered as a \dlydiff\ analogue of the BSQ equation.

For convenience, let us change notations of parameters appearing in the delay \kdv\ (\ref{dlykdv2}), its solution (\ref{dlykdv_sol}), and its nonlinear form (\ref{dlykdv_nl}) as follows:
\begin{align*}
    a_1\rightarrow a_3\,,\quad
    a_2\rightarrow a_4\,,\quad
    a_3\rightarrow a_1+a_2\,,\quad
    s\rightarrow y\,,\quad
    \sig\rightarrow\eta\,,\quad
    p_i\rightarrow-\frac{1}{3q_i}\,,\quad
    q_i\rightarrow-\frac{1}{3p_i}\,.
\end{align*}
Then we obtain the following bilinear equation from (\ref{dlykdv2}):
\begin{align}
    \label{dlybsq2}
    \left((a_1+a_2)D_x(a_3D_x+a_4D_y)
    +3\sinh^2(\xi D_x+\eta D_y)\right)
    f(x,y)\cdot f(x,y)=0\,.
\end{align}
Similarly, we obtain the following $N$-soliton solution from (\ref{dlykdv_sol}):
\begin{align}
    \label{dlybsq_sol}
    \fl f(x,y)
    =\det\left(\delta_{ij}+\frac{\Phi_j}{p_i-q_j}\right)_{1\leq i,j\leq N}
    =\sum_{I\subset\{1,\ldots,N\}}
    \prod_{i\in I}\Phi_i
    \prod_{i<j,\ i,j\in I}\frac{(p_i-p_j)(q_i-q_j)}{(p_i-q_j)(q_i-p_j)}\,,\\
    \fl\Phi_i
    =\mu_i\exp\left\{
    \frac{1}{2}\ \frac{a_4x-a_3y}{a_4\xi-a_3\eta}\log\bfrac{q_i}{p_i}
    +\frac{1}{2}\ \frac{-\eta x+\xi y}{a_4\xi-a_3\eta}\left(\frac{1}{q_i}-\frac{1}{p_i}\right)
    \right\}\,,\nonumber\\
    \fl0
    =(a_1+a_2)a_4\log\bfrac{q_i}{p_i}
    -3(a_4\xi-a_3\eta)(q_i-p_i)
    -(a_1+a_2)\eta\left(\frac{1}{q_i}-\frac{1}{p_i}\right)\,.\nonumber
\end{align}
Similarly, we obtain the following nonlinear equation from (\ref{dlykdv_nl}):
\begin{align}
    \label{dlybsq_nl}
    \fl (a_1+a_2)a_3\pd[2]{}{x}\log(1+b_1U(x,y))
    +(a_1+a_2)a_4\frac{\partial^2}{\partial x\partial y}\log(1+b_1U(x,y))\nonumber\\
    \fl \hspace{20mm}+\frac{3b_1}{4}\left\{U(x+2\xi,y+2\eta)+U(x-2\xi,y-2\eta)
    -2U(x,y)\right\}
    =0\,.
\end{align}
We claim that equation (\ref{dlybsq2}), which is equivalent to the delay \kdv, is the bilinear form of the delay BSQ equation.
In addition equation (\ref{dlybsq_sol}) is its $N$-soliton solution, and (\ref{dlybsq_nl}) is its nonlinear form.

To calculate the continuum limit of (\ref{dlybsq2}), (\ref{dlybsq_sol}), and (\ref{dlybsq_nl}), we set the parameters as follows:
\begin{align}
    \label{dlybsq_para}
    a_1=3\xi\,,\qquad
    a_2=4\xi^3\,,\qquad
    a_3=-\xi\,,\qquad
    a_4=-2\xi^2\,,\qquad
    \eta=\xi^2\,.
\end{align}
In addition we replace $p_i$ and $q_i$ as follows:
\begin{align}
    \label{dlybsq_rep}
    p_i\rightarrow\frac{1}{1+2\xi p_i}\,,\qquad
    q_i\rightarrow\frac{1}{1+2\xi q_i}\,.
\end{align}
Calculating the limit of (\ref{dlybsq2}) as $\xi\to0$, we obtain the bilinear form of the BSQ equation~\cite{Hirota-BSQ,Nimmo-BSQ}
\begin{align*}
    \left(-4D^2_x+3D^2_y+D^4_x\right)
    f(x,y)\cdot f(x,y)=0\,.
\end{align*}
By applying the limit $\xi\to0$ to (\ref{dlybsq_sol}), we can obtain the $N$-soliton solution of the BSQ equation~\cite{Hirota-BSQ,Nimmo-BSQ}
\begin{align*}
    \fl f(x,y)
    =\det\left(\delta_{ij}+\frac{\Phi_j}{p_i-q_j}\right)_{1\leq i,j\leq N}
    =\sum_{I\subset\{1,\ldots,N\}}
    \prod_{i\in I}\Phi_i
    \prod_{i<j,\ i,j\in I}\frac{(p_i-p_j)(q_i-q_j)}{(p_i-q_j)(q_i-p_j)}\,,\\
    \fl\Phi_i
    =\mu_ie^{(p_i-q_i)x+(p_i^2-q_i^2)y}\,,\qquad p_i-q_i=p_i^3-q_i^3\,.\nonumber
\end{align*}
Finally, putting $b_1=2\xi^2$, the limit of (\ref{dlybsq_nl}) as $\xi\to0$ is the nonlinear form of the BSQ equation~\cite{Hirota-BSQ,Nimmo-BSQ}
\begin{align*}
    -4U_{xx}+3U_{yy}+3(U^2)_{xx}+U_{xxxx}=0\,.
\end{align*}

\begin{rem}
We can also derive the delay BSQ equation (\ref{dlybsq2}) by the method in~\cite{Nakata1}.
First, the discrete BSQ equation introduced by Maruno and Kajiwara~\cite{Maruno-discrete_BSQ} is given by
\begin{align*}
    \del f_{n+1}^{m+1}f_{n}^{m}-(1+\del)f_{n}^{m+2}f_{n+1}^{m-1}+f_{n}^{m-1}f_{n+1}^{m+2}=0.
\end{align*}
This equation can be derived by applying the reduction condition
\begin{align*}
    f_{n,m,k+1}=f_{n,m-2,k}
\end{align*}
to the discrete \kp\ (\ref{dkp}) and setting $\del=a(b-c)/(c(a-b))$.
Now, we generalize the above reduction condition to
\begin{align*}
    f_{n,m,k+1}=f_{n-2\al,m-2-2\mu,k}\,,
\end{align*}
where the delay parameters $\al$ and $\mu$ are real constants.
Applying this reduction to the discrete \kp\ (\ref{dkp}) and setting $\eps\del=a(b-c)/(c(a-b))$, we obtain
\begin{align}
    \label{dlydisbsq}
    \eps\del f_{n+1+\al}^{m+1+\mu}f_{n-\al}^{m-\mu}-(1+\eps\del)f_{n+\al}^{m+2+\mu}f_{n+1-\al}^{m-1-\mu}+f_{n-\al}^{m-1-\mu}f_{n+1+\al}^{m+2+\mu}=0\,.
\end{align}
We claim that the bilinear equation (\ref{dlydisbsq}) is the delay discrete BSQ equation.
Now, applying the continuum limit
\begin{align*}
    \del\to0\,,\qquad
    n\del=r\,,\qquad
    \al\del=\rho=\const
\end{align*}
to the delay discrete BSQ equation (\ref{dlydisbsq}), we obtain
\begin{align}
    \label{dlysdbsq}
    \fl D_{r}f^{m+2+\mu}(r+\rho)\cdot f^{m-1-\mu}(r-\rho)\nonumber\\
    \fl \hspace{20mm}+\eps(f^{m+1+\mu}(r+\rho)f^{m-\mu}(r-\rho)-f^{m+2+\mu}(r+\rho)f^{m-1-\mu}(r-\rho))
    =0\,.
\end{align}
We can consider this bilinear equation (\ref{dlysdbsq}) is the delay semi-discrete BSQ equation.
(\ref{dlysdbsq}) can be rewritten as
\begin{align}
    \label{dlysdbsq2}
    \fl\left(\sinh\left(\frac{3D_m}{2}+\mu D_m+\rho D_r\right)D_r
    -2\eps\sinh\bfrac{D_m}{2}\sinh\left(D_m+\mu D_m+\rho D_r\right)\right)
    f_{n}^{k}(t)\cdot f_{n}^{k}(t)\nonumber\\
    \fl \hspace{20mm}=0\,.
\end{align}
Finally, applying the continuum limit
\begin{align*}
    \fl \eps\to0\,,\quad
    D_r=\frac{2}{3}(a_1+a_2)D_x\,,\\
    \fl D_m=2(\xi D_x+\eta D_y)+\frac{2\eps}{3}(a_3D_x+a_4D_y)\,,\\
    \fl \mu D_m+\rho D_r=-3(\xi D_x+\eta D_y)
\end{align*}
to the delay semi-discrete BSQ equation (\ref{dlysdbsq2}), we obtain the delay BSQ equation (\ref{dlybsq2}).
Note that this continuum limit is obtained in the same way as Remark \ref{rem_kdv1}.
\end{rem}

\end{subsection}

\end{section}

\begin{section}{An integrable delay \kp}
\label{sec_kp}

Although we constructed the delay KdV and delay BSQ equations in section \ref{sec_kdv}, we have not yet presented the soliton equation which reduces to them.
In this section, we first construct the delay \kp\ and its Casorati determinant solution by a transformation from the 2DTL.
Then we derive again the delay KdV and delay BSQ equations and their $N$-soliton solutions by reductions of the delay \kp.

\begin{subsection}{The delay \kp}

We start from the bilinear form of the 2DTL~\cite{Hirota-d2DTL,Hirota-direct}
\begin{align}
    \label{2dtl}
    \left(D_tD_z
    -4\sinh^2\bfrac{D_n}{2}\right)
    f_{n}(t,z)\cdot f_{n}(t,z)=0\,.
\end{align}
The Casorati (Wronski) determinant solution~\cite{Hirota-d2DTL,Hirota-direct} of the 2DTL (\ref{2dtl}) is given as follows:
\begin{align}
    \label{2dtl_cas}
    \fl f_{n}(t,z)
    =\left|
    \begin{array}{cccc}
        \phi_1(0;n,t,z) & \phi_1(1;n,t,z) & \cdots & \phi_1(N-1;n,t,z)\\
        \phi_2(0;n,t,z) & \phi_2(1;n,t,z) & \cdots & \phi_2(N-1;n,t,z)\\
        \vdots & \vdots & \cdots & \vdots \\
        \phi_N(0;n,t,z) & \phi_N(1;n,t,z) & \cdots & \phi_N(N-1;n,t,z)
    \end{array}
    \right|\,,\\
    \fl\phi_i(j+1;n,t,z)=\phi_i(j;n+1,t,z)\,,\nonumber\\
    \fl\phi_i(j;n+1,t,z)=\pd{}{t}\phi_i(j;n,t,z)\,,\nonumber\\
    \fl\phi_i(j;n-1,t,z)=-\pd{}{z}\phi_i(j;n,t,z)\,.\nonumber
\end{align}

\begin{rem}
The $N$-soliton solution of the 2DTL is given by setting the elements in (\ref{2dtl_cas}) as follows:
\begin{align}
    \label{2dtl_sol}
    \phi_i(j;n,t,z)
    =\be_ip_i^{j+n}\exp\left(p_it-\frac{1}{p_i}z\right)
    +\gam_iq_i^{j+n}\exp\left(q_it-\frac{1}{q_i}z\right)\,,
\end{align}
where $\be_i$ and $\gam_i$ are real constants.
\end{rem}

Let us apply the following transformation to the 2DTL (\ref{2dtl}) and its solution (\ref{2dtl_cas}):
\begin{align}
    \label{2dtl_vartrans}
    \begin{aligned}
    \fl n=\frac{a_2a_4x-a_1a_4s-a_2a_3y}
    {2a_2a_4\xi-2a_1a_4\sig-2a_2a_3\eta}\,,\\
    \fl t=\frac{3a_4(-\sig x+\xi s)+3a_3(-\eta s+\sig y)}
    {2a_2a_4\xi-2a_1a_4\sig-2a_2a_3\eta}\,,\qquad
    z=-\frac{a_2(-\eta x+\xi y)+a_1(\eta s-\sig y)}
    {2a_2a_4\xi-2a_1a_4\sig-2a_2a_3\eta}\,,
    \end{aligned}
\end{align}
where $x,s,y$ are continuous variables, and the parameters $a_1,a_2,a_3,a_4,\xi,\sig,\eta$ are real constants.
The relations (\ref{2dtl_vartrans}) lead to the following ones:
\begin{align*}
    D_{n}=2\xi D_x+2\sig D_s+2\eta D_y\,,\quad
    D_{t}=\frac{2}{3}(a_1D_x+a_2D_s)\,,\quad
    D_{z}=-2(a_3D_x+a_4D_y)\,.
\end{align*}
Substituting these relations into the 2DTL (\ref{2dtl}), we obtain
\begin{align}
    \label{dlykp2}
    \left((a_1D_x+a_2D_s)(a_3D_x+a_4D_y)
    +3\sinh^2(\xi D_x+\sig D_s+\eta D_y)\right)
    f\cdot f=0\,.
\end{align}
This bilinear equation is equivalent to
\begin{align}
    \label{dlykp1}
    \fl (a_1D_x+a_2D_s)(a_3D_x+a_4D_y)f\cdot f
    +\frac{3}{2}\left(\ov{f}\un{f}-ff\right)
    =0\,,\\
    \fl \ov{f}\equiv f(x+2\xi,s+2\sig,y+2\eta)\,,\quad
    \un{f}\equiv f(x-2\xi,s-2\sig,y-2\eta)\,.\nonumber
\end{align}
We claim that the bilinear equation (\ref{dlykp1}) (or (\ref{dlykp2})) is the delay \kp.
Next, applying the transformation (\ref{2dtl_vartrans}) to (\ref{2dtl_cas}), we obtain the Casorati determinant solution of the delay \kp\ as follows:
\begin{align}
    \label{dlykp_cas}
    \fl f(x,s,y)
    =\left|
    \begin{array}{cccc}
        \phi_1(0;x,s,y) & \phi_1(1;x,s,y) & \cdots & \phi_1(N-1;x,s,y)\\
        \phi_2(0;x,s,y) & \phi_2(1;x,s,y) & \cdots & \phi_2(N-1;x,s,y)\\
        \vdots & \vdots & \cdots & \vdots \\
        \phi_N(0;x,s,y) & \phi_N(1;x,s,y) & \cdots & \phi_N(N-1;x,s,y)
    \end{array}
    \right|\,,\\
    \fl\phi_i(j+1;x,s,y)=\ov{\phi_i}(j;x,s,y)\,,\nonumber\\
    \fl\ov{\phi_i}(j;x,s,y)=\frac{2}{3}\left(a_1\pd{}{x}+a_2\pd{}{s}\right)\phi_i(j;x,s,y)\,,\nonumber\\
    \fl\un{\phi_i}(j;x,s,y)=2\left(a_3\pd{}{x}+a_4\pd{}{y}\right)\phi_i(j;x,s,y)\,,\nonumber
\end{align}
where
\begin{align*}
    \fl \ov{\phi_i}(j;x,s,y)\equiv\phi_i(j;x+2\xi,s+2\sig,y+2\eta)\,,\\
    \fl \un{\phi_i}(j;x,s,y)\equiv\phi_i(j;x-2\xi,s-2\sig,y-2\eta)\,.
\end{align*}

\begin{rem}
Through the transformation (\ref{2dtl_vartrans}), the function (\ref{2dtl_sol}) is transformed into
\begin{align*}
    \fl \begin{aligned}\phi_i(j;x,s,y)
    &=\be_ip_i^{j}\exp(P(p_i)x+Q(p_i)s+R(p_i)y)\nonumber\\
    &\quad+\gam_iq_i^{j}\exp(P(q_i)x+Q(q_i)s+R(q_i)y)\,,\end{aligned}\\
    \fl P(p)
    \equiv\frac{a_2a_4\log p
    -3a_4\sig p
    -a_2\eta(1/p)}
    {2a_2a_4\xi-2a_1a_4\sig-2a_2a_3\eta}\,,\nonumber\\
    \fl Q(p)
    \equiv\frac{-a_1a_4\log p
    +3(a_4\xi-a_3\eta)p
    +a_1\eta(1/p)}
    {2a_2a_4\xi-2a_1a_4\sig-2a_2a_3\eta}\,,\nonumber\\
    \fl R(p)
    \equiv\frac{-a_2a_3\log p
    +3a_3\sig p
    +(a_2\xi-a_1\sig)(1/p)}
    {2a_2a_4\xi-2a_1a_4\sig-2a_2a_3\eta}\,.\nonumber
\end{align*}
Thus the $N$-soliton solution of the delay \kp\ is obtained by setting $\phi_i(j;x,s,y)$ as above.
Note that the $N$-soliton solution in the Gram determinant form is given as follows:
\begin{align}
    \label{dlykp_sol}
    \fl f(x,s,y)
    =\det\left(\delta_{ij}+\frac{\Phi_j}{p_i-q_j}\right)_{1\leq i,j\leq N}
    =\sum_{I\subset\{1,\ldots,N\}}
    \prod_{i\in I}\Phi_i
    \prod_{i<j,\ i,j\in I}\frac{(p_i-p_j)(q_i-q_j)}{(p_i-q_j)(q_i-p_j)}\,,\\
    \fl\Phi_i
    =\mu_i\exp\left\{P(p_i,q_i)x+Q(p_i,q_i)s+R(p_i,q_i)y\right\}\,,\nonumber\\
    \fl P(p_i,q_i)
    \equiv\frac{a_2a_4\log(q_i/p_i)
    -3a_4\sig(q_i-p_i)
    -a_2\eta\left(1/q_i-1/p_i\right)}
    {2a_2a_4\xi-2a_1a_4\sig-2a_2a_3\eta}\,,\nonumber\\
    \fl Q(p_i,q_i)
    \equiv\frac{-a_1a_4\log(q_i/p_i)
    +3(a_4\xi-a_3\eta)(q_i-p_i)
    +a_1\eta\left(1/q_i-1/p_i\right)}
    {2a_2a_4\xi-2a_1a_4\sig-2a_2a_3\eta}\,,\nonumber\\
    \fl R(p_i,q_i)
    \equiv\frac{-a_2a_3\log(q_i/p_i)
    +3a_3\sig(q_i-p_i)
    +(a_2\xi-a_1\sig)\left(1/q_i-1/p_i\right)}
    {2a_2a_4\xi-2a_1a_4\sig-2a_2a_3\eta}\,,\nonumber
\end{align}
where $\mu_i$ are real constants.
These $P,Q,R$ satisfy the dispersion relation
\begin{align*}
    (a_1P+a_2Q)(a_3P+a_4R)
    +3\sinh^2(\xi P+\sig Q+\eta R)
    =0\,.
\end{align*}
\end{rem}

Next, we show the delay \kp\ has a continuum limit to the \kp.
We first set the parameters as follows:
\begin{align}
    \label{dlykp_para}
    a_1=3\xi\,,\quad
    a_2=4\xi^3\,,\quad
    a_3=-\xi\,,\quad
    a_4=-2\xi^2\,,\quad
    \sig=0\,,\quad
    \eta=\xi^2\,.
\end{align}
Note that this setting is consistent with (\ref{dlykdv_para}) and (\ref{dlybsq_para}).
In addition we apply the replacement
\begin{align*}
    p_i\rightarrow\frac{1}{1+2\xi p_i}\,,\qquad
    q_i\rightarrow\frac{1}{1+2\xi q_i}\,.
\end{align*}
Note that this replacement is exactly the same as (\ref{dlykdv_rep}) and (\ref{dlybsq_rep}).
Now, calculating the limit of the delay \kp\ (\ref{dlykp2}) as $\xi\to0$, we obtain the bilinear form of the \kp~\cite{Hirota-direct}
\begin{align*}
    \left(-4D_sD_x+3D^2_y+D^4_x\right)
    f\cdot f=0\,.
\end{align*}
In addition, by applying the limit $\xi\to0$ to (\ref{dlykp_sol}), we can obtain the $N$-soliton solution of the \kp~\cite{Hirota-direct}
\begin{align*}
    \fl f(x,y,s)
    =\det\left(\delta_{ij}+\frac{\Phi_j}{p_i-q_j}\right)_{1\leq i,j\leq N}
    =\sum_{I\subset\{1,\ldots,N\}}
    \prod_{i\in I}\Phi_i
    \prod_{i<j,\ i,j\in I}\frac{(p_i-p_j)(q_i-q_j)}{(p_i-q_j)(q_i-p_j)}\,,\\
    \fl\Phi_i
    =\mu_ie^{(p_i-q_i)x+(p_i^2-q_i^2)y+(p_i^3-q_i^3)s}\,.\nonumber
\end{align*}

Finally, we present the nonlinear form of the delay \kp\ via the dependent variable transformation
\begin{align}
    \label{dlykp_trans}
    1+b_1U(x,s,y)=\frac{f(x+2\xi,s+2\sig,y+2\eta)f(x-2\xi,s-2\sig,y-2\eta)}{f(x,s,y)f(x,s,y)}\,,
\end{align}
where $b_1$ is a real constant.
By using this, bilinear equation (\ref{dlykp1}) is transformed into the following \dlydiff\ equation:
\begin{align}
    \label{dlykp_nl}
    \fl a_1a_3\pd[2]{}{x}\log(1+b_1U)
    +a_2a_3\frac{\partial^2}{\partial s\partial x}\log(1+b_1U)\nonumber\\
    \fl \hspace{20mm}+a_1a_4\frac{\partial^2}{\partial x\partial y}\log(1+b_1U)
    +a_2a_4\frac{\partial^2}{\partial s\partial y}\log(1+b_1U)\nonumber\\
    \fl \hspace{20mm}+\frac{3b_1}{4}\left(\ov{U}+\un{U}-2U\right)
    =0\,,\\
    \fl \ov{U}\equiv U(x+2\xi,s+2\sig,y+2\eta)\,,\quad
    \un{U}\equiv U(x-2\xi,s-2\sig,y-2\eta)\,.\nonumber
\end{align}
(\ref{dlykp_nl}) is the nonlinear form of the delay \kp.
Under the conditions (\ref{dlykp_para}) and $b_1=2\xi^2$, the limit of (\ref{dlykp_nl}) as $\xi\to0$ is the nonlinear form of the \kp~\cite{Hirota-direct}
\begin{align*}
    \left(-4U_s+6UU_x+U_{xxx}\right)_x+3U_{yy}=0\,.
\end{align*}

\begin{rem}
Transformation (\ref{dlykp_trans}) is understood in the same way as Remark \ref{rem_kdv2}.
When $b_1=2\xi^2$ and $\xi\to0$, (\ref{dlykp_trans}) becomes
\begin{align*}
    U(x,s,y)=2(\log f(x,s,y))_{xx}\,,
\end{align*}
which is the transformation between the bilinear and nonlinear KP equations.
\end{rem}

\end{subsection}

\begin{subsection}{Reductions of the delay \kp}

Let us consider reductions of the delay \kp\ (\ref{dlykp2}) and its $N$-soliton solution (\ref{dlykp_sol}).

First of all, note that the \kdv\ and BSQ equation can be derived by reductions of the \kp, which is given by
\begin{align*}
    \left(-4D_sD_x+3D^2_y+D^4_x\right)
    f(x,y,s)\cdot f(x,y,s)=0\,.
\end{align*}
Applying the reduction condition $f_y(x,y,s)=0$ to the \kp, we actually obtain the \kdv
\begin{align*}
    \left(-4D_sD_x+D^4_x\right)
    f(x,s)\cdot f(x,s)=0\,.
\end{align*}
On the other hand, applying the reduction condition $f_x(x,y,s)=f_s(x,y,s)$ to the \kp, we obtain the BSQ equation
\begin{align*}
    \left(-4D^2_x+3D^2_y+D^4_x\right)
    f(x,y)\cdot f(x,y)=0\,.
\end{align*}
Inspired by these derivations, we propose two reductions of the delay \kp.

First, we apply the reduction condition
\begin{align}
    \label{reduction1}
    \pd{f}{y}=0\,,\qquad
    \eta=0
\end{align}
to the delay \kp\ (\ref{dlykp2}).
We can easily check that (\ref{dlykp2}) reduces to the delay \kdv\ (\ref{dlykdv2}).
To realize the reduction condition (\ref{reduction1}) for the $N$-soliton solution, we set
\begin{align*}
    R(p_i,q_i)=0\,,\qquad
    \eta=0
\end{align*}
on the solution of the delay \kp\ (\ref{dlykp_sol}).
The constraint $R(p_i,q_i)=0$ is equivalent to the dispersion relation of the delay \kdv, which is the last equation of (\ref{dlykdv_sol}).
Therefore, solution (\ref{dlykp_sol}) under $R(p_i,q_i)=0$ and $\eta=0$ reduces to the $N$-soliton solution of the delay \kdv\ (\ref{dlykdv_sol}).

Next, we apply the reduction condition
\begin{align}
    \label{reduction2}
    \pd{f}{x}=\pd{f}{s}\,,\qquad
    \sig=0
\end{align}
to the delay \kp\ (\ref{dlykp2}).
We can easily check that (\ref{dlykp2}) reduces to the delay BSQ equation (\ref{dlybsq2}).
To realize the reduction condition (\ref{reduction2}) for the $N$-soliton solution, we set
\begin{align*}
    P(p_i,q_i)=Q(p_i,q_i)\,,\qquad
    \sig=0
\end{align*}
on the solution of the delay \kp\ (\ref{dlykp_sol}).
Similar to the above example, solution (\ref{dlykp_sol}) reduces to the $N$-soliton solution of the delay BSQ equation (\ref{dlybsq_sol}).

In the above constructions of the delay KdV and delay BSQ equations, we did not need formal continuum limits. 

\end{subsection}

\end{section}

\begin{section}{Conclusions}
\label{sec_con}

We have constructed \dlydiff\ analogues of the KdV, BSQ, and KP equations with their $N$-soliton solutions.
In the small limit of the delay parameters, these delay soliton equations reduced to the KdV, BSQ, and KP equations respectively.
The delay BSQ equation is equivalent to the delay \kdv, and the delay \kp\ is equivalent to the 2DTL.
In the process of their construction, we also obtained delay-difference analogues of the KdV and BSQ equations and their $N$-soliton solutions.
In addition, by reductions of the delay \kp, we derived again the delay KdV and delay BSQ equations and their $N$-soliton solutions.
Now, we can clearly display the relationship of the KdV and KP families in Fig.~\ref{fig_KdV}.

\begin{figure*}
    \includegraphics
    {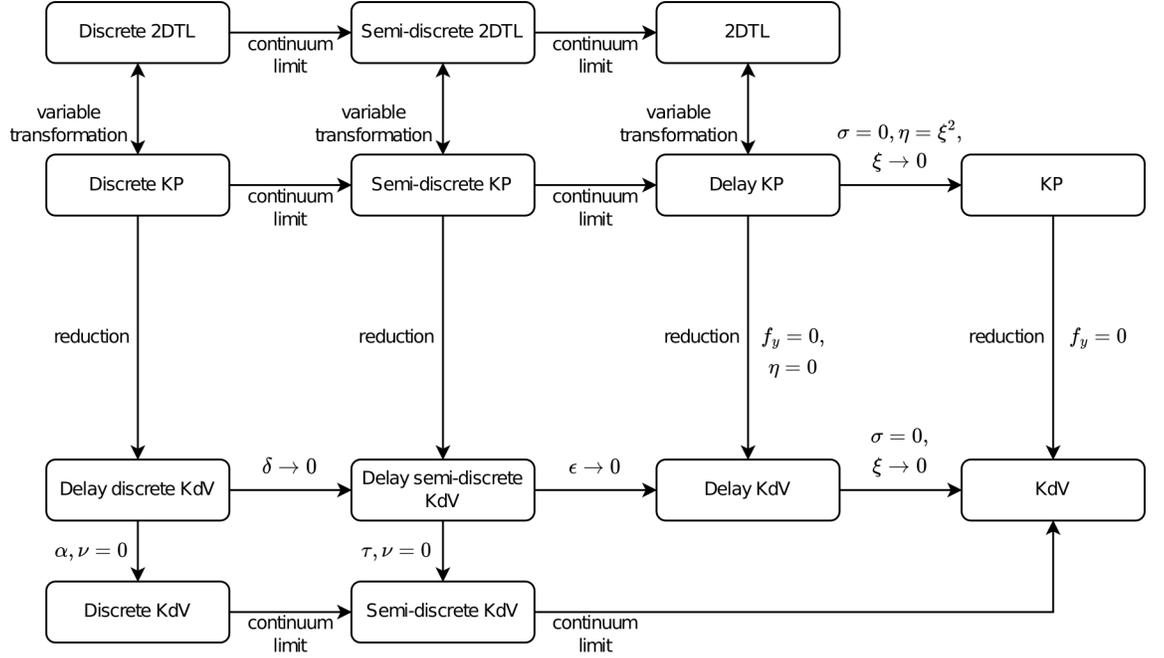}
    \caption{\label{fig_KdV}
    Relationship diagram of the KdV and KP families.}
\end{figure*}

The derivations of the delay LV, delay TL, and delay sG equations were perfectly parallel to the non-delay version~\cite{Nakata1}.
However, some delay soliton equations (such as the delay \kdv) cannot be obtained by this idea.
In this paper, the delay \kdv\ succeeded to be obtained by a technique which is not parallel to the non-delay version (as we discussed in Remark \ref{rem_kdv1}).

This paper and our previous studies~\cite{Tsunematsu,Nakata1} mainly focus on proposing delay soliton equations and deriving their soliton solutions.
Thus we have not yet revealed many things about them such as the Lax pairs, ultra-discretization, singularity confinement, and relationships to the \dlydiff\ \pain\ equations.
In particular, it should be investigated whether delay \pain\ equations are derived by reductions of delay soliton equations.
In addition, it remains unclear if the delay soliton equations can model and analyze feedback systems such as traffic flow.
We intend to address these problems in future studies.

\end{section}

\begin{section}*{Acknowledgments}
The author would like to thank Prof.~Maruno, Waseda university, for stimulating conversations and his helpful comments on this work.
\end{section}


\begin{thebibliography}{0}


\bibitem{nakanishi}
K. Nakanishi, K. Itoh, Y. Igarashi and M. Bando,
\textit{Phys. Rev. E} {\bf 55}, 6519 (1997)

\bibitem{hasebe}
K. Hasebe, A. Nakayama and Y. Sugiyama,
\textit{Phys. Lett. A} {\bf 259}, 135 (1999)

\bibitem{kanai}
Y. Tutiya and M. Kanai,
\textit{J. Phys. Soc. Jpn.} {\bf 76}, 083002 (2007)


\bibitem{Quispel}
G. R. W. Quispel, H. W. Capel and R. Sahadevan,
\textit{Phys. Lett. A} {\bf 170}, 379 (1992)

\bibitem{Ramani1}
A. Ramani, B. Grammaticos and K. M. Tamizhmani,
\textit{J. Phys. A: Math. Gen.} {\bf 25}, L883 (1992)

\bibitem{Ramani2}
A. Ramani, B. Grammaticos and K. M. Tamizhmani,
\textit{J. Phys. A: Math. Gen.} {\bf 26}, L53 (1993)

\bibitem{Gram}
B. Grammaticos, A. Ramani and I. C. Moreira,
\textit{Physica A} {\bf 196}, 574 (1993)

\bibitem{Levi}
D. Levi and P. Winternitz,
\textit{J. Math. Phys.} {\bf 34}, 3713 (1993)

\bibitem{Joshi1}
N. Joshi,
\textit{J. Phys. A: Math. Theor.} {\bf 42}, 022001 (2008)

\bibitem{Joshi2}
N. Joshi and P. E. Spicer,
\textit{J. Phys. Soc. Jpn.} {\bf 78}, 094006 (2009)

\bibitem{Carstea}
A. S. Carstea,
\textit{J. Phys. A: Math. Theor.} {\bf 44}, 105202 (2011)

\bibitem{Viallet}
C. M. Viallet,
(arXiv:1408.6161) (2014)

\bibitem{Halburd}
R. Halburd and R. Korhonen,
\textit{Proc. Am. Math. Soc.} {\bf 145}, 2513 (2017)

\bibitem{Berntson}
B. K. Berntson,
\textit{SIGMA} {\bf 14}, 020 (2018)

\bibitem{Stokes}
A. Stokes,
\textit{J. Phys. A: Math. Theor.} {\bf 53}, 435201 (2020)

\bibitem{Ablowitz}
J. Villarroel and M. J. Ablowitz,
\textit{Phys. Lett. A} {\bf 180}, 413 (1993)

\bibitem{Tsunematsu}
A. Tsunematsu, K. Nakata, Y. Tanaka and K. Maruno,
(arXiv:2201.09473) (2021)

\bibitem{Nakata1}
K. Nakata and K. Maruno,
\textit{J. Phys. A: Math. Theor.} {\bf 55}, 335201 (2022)


\bibitem{Hirota-discrete_KP}
R. Hirota,
\textit{J. Phys. Soc. Jpn.} {\bf 50}, 3785 (1981)

\bibitem{Miwa-discrete_KP}
T. Miwa,
\textit{Proc. Japan Acad. A} {\bf 58}, 9 (1982)

\bibitem{Ohta-discrete_KP}
Y. Ohta, R. Hirota, S. Tsujimoto and T. Imai,
\textit{J. Phys. Soc. Jpn.} {\bf 62}, 1872 (1993)

\bibitem{Hirota-discrete_KdV}
R. Hirota,
\textit{J. Phys. Soc. Jpn.} {\bf 43}, 1424 (1997)

\bibitem{Maruno-discrete_soliton_equations}
K. Maruno, K. Kajiwara, S. Nakao and M. Oikawa,
\textit{Phys. Lett. A} {\bf 229}, 173 (1997)

\bibitem{Li-semi_discrete_KP}
C. X. Li, S. Lafortune and S. F. Shen,
\textit{J. Math. Phys.} {\bf 57}, 053503 (2016)

\bibitem{Hirota-KdV}
R. Hirota,
\textit{Phys. Rev. Lett} {\bf 27}, 1192 (1971)

\bibitem{Hirota-direct}
R. Hirota,
\textit{The Direct Method in Soliton Theory}
(Cambridge University Press, 2004)

\bibitem{Hirota-BSQ}
R. Hirota and J. Satsuma,
\textit{Prog. Theor. Phys.} {\bf 57}, 797 (1977)

\bibitem{Nimmo-BSQ}
J. J. C. Nimmo and N. C. Freeman,
\textit{Phys. Lett. A} {\bf 95}, 4 (1983)

\bibitem{Maruno-discrete_BSQ}
K. Maruno and K. Kajiwara,
\textit{Appl. Anal.} {\bf 89}, 593 (2010)

\bibitem{Hirota-d2DTL}
R. Hirota, M. Ito and F. Kako,
\textit{Prog. Theor. Phys. Suppl.} {\bf 94}, 42 (1988)

\end{thebibliography}
\end{document}